\documentclass[aps,prl,superscriptaddress,twocolumn,showpacs]{revtex4-1}

\usepackage{graphicx}
\usepackage{psfig}
\usepackage{amsmath}
\usepackage{caption2}
\usepackage{floatflt}
\usepackage{mathrsfs}
\usepackage{color}

\begin{document}
\title{Detuning-controlled internal oscillations in an exciton-polariton condensate}
\author{\firstname{N.~S.}~\surname{Voronova}}
\email{nsvoronova@mephi.ru}
\affiliation{National Research Nuclear University MEPhI (Moscow Engineering Physics Institute), 115409 Moscow, Russia}
\author{\firstname{A.~A.}~\surname{Elistratov}}
\affiliation{ Institute for Nanotechnology in Microelectronics RAS, 119334 Leninskiy ave. 32a, Moscow, Russia}
\author{\firstname{Yu.~E.}~\surname{Lozovik}}
\affiliation{Institute for Spectroscopy RAS, 142190 Troitsk, Moscow, Russia}
\affiliation{Moscow Institute of Physics and Technology (State University), 141700 Dolgoprudny, Moscow region, Russia}
\affiliation{Moscow Institute of Electronics and Mathematics, HSE, 101000 Moscow, Russia}

\begin{abstract}
We theoretically analyze exciton-photon oscillatory dynamics within a homogenous polariton gas in presence of energy detuning between the cavity and quantum well modes. Whereas pure Rabi oscillations consist of the particle exchange between the photon and excitons states in the polariton system without any oscillations of the phases of the two sub-condensates, we demonstrate that any non-zero detuning results in oscillations of the relative phase of the photon and exciton macroscopic wave functions. Different initial conditions reveal a variety of behaviors of the relative phase between the two condensates, and a crossover from Rabi-like to Josephson-like oscillations is predicted.
\end{abstract}

\maketitle

Exciton-polaritons are new mixed eigenmodes resulting from strong coupling between the photon state in a microcavity and exciton state in a quantum well, which inherit properties of both light and matter \cite{MCpolaritons}. Polaritons interact without dephasing due to elastic collisions in their excitonic component, while the photon component provides them with extremely light effective mass allowing Bose-Einstein condensation (BEC) at high critical temperatures \cite{richard,kasprzak,balili}. In the recent years, great wealth of experimental \cite{lagoudakis,sanvitto_v,amo1,amo2,bloch,amo3,lagoudakis1,abbarchi,Sven} and theoretical \cite{szymanska,marchetti,flayac,solnyshkov,carusotto,ciuti,kavokin1,shelykh, kammann,voronova,wouters,shelykh1,sarchi,borgh,read} works in the field of polariton BEC demonstrated the coherent effects analogous to those in atomic condensates \cite{giovanazzi,albiez,levy,leggett2}, superconductors \cite{anderson2,backhaus}, or liquid Helium \cite{anderson,leggett,wheatley}. At the same time, due to their non-trivial dispersion, short lifetime, and therefore non-equilibrium nature of condensation, polariton quantum properties \cite{QFL} differ considerably from those appearing in BECs of atoms or in superconducting systems. The phenomena investigated include condensation in traps \cite{balili,bloch}, superfluidity \cite{carusotto,amo1,amo2,amo3}, vortices \cite{lagoudakis,sanvitto_v,marchetti,voronova}, solitons \cite{flayac,solnyshkov,amo3}, polariton polarization features \cite{kavokin1,shelykh,kammann}, and Josephson phenomena \cite{wouters,shelykh1,sarchi,borgh,read,lagoudakis1,abbarchi,Sven}. Latest experiments to date \cite{laussy,sanvitto1} report observation and fine control of such inherent features of polariton systems as relaxation oscillations and Rabi oscillations. There has as well been some theoretical effort dedicated to polariton Rabi dynamics, revealing possible ways to increase the coherence time and proposing qubits and logic gates based on exciton-polaritons \cite{glazov,liew}, and stochastic processes within polariton condensates \cite{sheremet}.

Strong coupling regime which is considered here is defined as the Rabi splitting between the upper (UP) and lower (LP) polariton branches at the anticrossing being large compared to their emission linewidths \cite{MCpolaritons}. The frequency of this two-level oscillator depends on the Rabi frequency $\Omega_R$ and the detuning between energies of bare photon and exciton states at zero wavevector. From the point of view of the initial exciton and photon states, if the detuning is zero, the polariton gas is a half-and-half mixture of the two coherent components constantly performing mutual transformations, hence one may say that Rabi oscillations are density oscillations between the `photon' and `exciton' condensates. The picture becomes more complicated when the photon and exciton dispersions are shifted with respect to each other: internal oscillations between exciton and photon components involve oscillations of the relative phase of the two macroscopic wave functions.

In this work, we present fully analytical investigation of internal oscillations in the two-component polariton system and discuss possible regimes of the dynamics. We consider an idealized polariton gas with constant chemical potential and neglect all non-equilibrium effects associated with particles gain and dissipation. For this model condensate, we demonstrate that at any non-zero detuning different types of oscillations are possible, from harmonic and anharmonic modifications of Rabi oscillations up to the transition to a so-called Josephson regime analogous to internal Josephson effect in a two-state BEC of $^{87}$Rb atoms \cite{matthews,matthews1}. We also address interaction-induced corrections to our analytical solutions brought into the system by possible increase of the polariton density.

Within the mean field approach \cite{ciuti}, temporal evolution of the macroscopic wave functions of cavity photons $\psi_C$ and quantum well excitons $\psi_X$ is described by the coupled Schr\"{o}dinger and Gross-Pitaevskii equations,
\begin{equation}\label{GPE_1}
i\hbar\partial_t\psi_C = \Bigl[E_C^0-\frac{\hbar^2\nabla^2}{2m_C}\Bigr]\psi_C + \frac{\hbar\Omega_R}{2}\,\psi_X,
\end{equation}
\begin{equation}\label{GPE_2}
i\hbar\partial_t\psi_X=\Bigl[E_X^0-\frac{\hbar^2\nabla^2}{2m_X}+\bar{g}|\psi_X|^2\Bigr]\psi_X + \frac{\hbar\Omega_R}{2}\,\psi_C,
\end{equation}
with $E_{C,X}^0$ the bottoms of energy dispersions and $m_{C,X}$ the effective masses of photons and excitons. $\bar{g}>0$ is the constant of repulsive exciton-exciton interaction. Particle transfer between the subsystems is described by the coupling term $\sim\hbar\Omega_R/2$, and we neglected any external potentials and spin degree of freedom.

When the system is in the strong coupling regime, the polariton state is an eigenstate with an equal (in the absence of interactions) superposition of a photon and an exciton. The positive sign chosen in Eqs.~(\ref{GPE_1}), (\ref{GPE_2}) in front of the coupling term $\hbar\Omega_R/2$ imposes that the antisymmetric mode $(\psi_C-\psi_X)/\sqrt{2}$ with the relative phase $\pi$ is the lower energy level (\textit{i.e.} corresponds to the LP state) while the symmetric mode $(\psi_C+\psi_X)/\sqrt{2}$ with zero relative phase is the upper one. An initial state of the polariton system, being some linear combination of these two modes, results in density oscillations between the photon and exciton subsystems. When interactions are present, the effective lower energy level is blueshifted (while the upper energy level appears redshifted), and the eigenmodes are no longer the antisymmetric and the symmetric ones. Still, the relative phase oscillations discussed below go around the time-average values $\pi$ or $0$.

Considering homogeneous case for simplicity and assuming momentum equal to zero, we neglect spatial derivatives in Eqs.~(\ref{GPE_1}) and (\ref{GPE_2}).
After the transformation $\psi_{C,X}(t)=\sqrt{n_{C,X}(t)}\,e^{iS_{C,X}(t)}$ we get four non-linear dynamical equations for photon and exciton populations $n_{C,X}(t)$ and phases $S_{C,X}(t)$:
\begin{equation}
\partial_t n_{C,X} = \mp\sqrt{n_C n_X}\,\sin\left(S_C-S_X\right),
\end{equation}
\begin{equation}\label{phase_phot}
\partial_t S_C = -\epsilon^0_C -\frac{1}{2}\sqrt{\frac{n_X}{n_C}}\,\cos\left(S_C-S_X\right),
\end{equation}
\begin{equation}\label{phase_exc}
\partial_t S_X = - \epsilon^0_X - gn_X - \frac{1}{2}\sqrt{\frac{n_C}{n_X}}\,\cos\left(S_C-S_X\right),
\end{equation}
where we have rescaled lengths and energies in terms of $\sqrt{\hbar/m_C\Omega_R}$ and $\hbar\Omega_R$, respectively, time as $t\Omega_R \rightarrow t$, and the wave functions as $\psi_{C,X}/\sqrt{\hbar/m_C\Omega_R}\rightarrow\psi_{C,X}$.

In order to investigate the dynamics, it is convenient to introduce new variables: relative phase $S(t)=S_C(t)-S_X(t)$ and population imbalance $\rho(t) = (n_C(t)-n_X(t))/n$, where $n=n_C(t)+n_X(t)$ is the total number of polaritons. Variables $\rho$ and $S$ obey the coupled equations
\begin{equation}\label{rho}
\dot{\rho} = -\sqrt{1-\rho^2}\,\sin S,
\end{equation}
\begin{equation}\label{s}
\dot{S} =  -\delta + \frac{gn}{2} -\frac{gn}{2}\,\rho + \frac{\rho}{\sqrt{1-\rho^2}}\,\cos S.
\end{equation}
The dimensionless detuning $\delta=\epsilon^0_C-\epsilon^0_X$ and the blueshift value $gn/2$ are the parameters which determine different regimes of the system behavior. For a system with constant chemical potential the equations (\ref{rho}), (\ref{s}) are Hamiltonian with the conserved energy $H(S,\rho) = (\delta-gn/2)\rho + gn\rho^2/4 + \sqrt{1-\rho^2}\,\cos S$, where total population $n$ is constant. Equations (\ref{rho}) and (\ref{s}) admit analytical solution in terms of quadratures:
\begin{equation}\label{quadra1}
\cos S = \frac{H-\frac{gn}{2}\,\frac{\rho^2}{2}-\left(\delta-\frac{gn}{2}\right)\rho}{\sqrt{1-\rho^2}},
\end{equation}
\begin{equation}\label{integral}
t = \mp\int\frac{d\rho}{\sqrt{1-\rho^2-\left(H-\frac{gn}{2}\,\frac{\rho^2}{2}-\left(\delta-\frac{gn}{2}\right)\rho\right)^2}}\,.
\end{equation}

After obtaining formal solution of the evolution equations, it is worth noting that the interaction constant $g$ is of the order of $10^{-3}$ (estimated from $\bar{g}=0.015$~meV$\cdot\mu$m$^2$ \cite{kasprzak}), while the unscaled $n$ is of the order of unity (which corresponds to $\sim10^{10}$~cm$^{-2}$ \cite{kasprzak}). Hence, for the closed conservative system the blueshift parameter $gn/2$ is always of the order of $10^{-3}$, and the effect of interactions on internal oscillations should be negligible. We estimate the upper limit for realistic polariton densities as $n_{cr}\sim 10^{11}$--$10^{12}$~cm$^{-2}$ \cite{kir}. In our numerical simulations, we raised the total density $n$ up to $0.5\times10^{12}$~cm$^{-2}$. However, as shown further, even for large densities neglecting the interactions results in little loss of accuracy.

In the absence of interactions, the integral in (\ref{integral}) reveals an explicit solution for the population imbalance:
\begin{equation}\label{rho_analyt}
\rho(t)= \frac{h\delta}{\omega^2}+\frac{\sqrt{\omega^2-h^2}}{\omega^2}\,\sin(\omega t-\varphi),
\end{equation}
where $h=\delta\rho(0)+\sqrt{1-\rho(0)^2}\cos S(0)=$~const is the energy (per particle) defined by the detuning and the initial conditions, $\omega=\sqrt{1+\delta^2}$ is the renormalized frequency of internal oscillations (in scaled units, it corresponds to $\Omega_R\sqrt{1+\delta^2}$), and $\varphi=\arcsin[(h\delta-\rho(0)\omega^2)/ \sqrt{\omega^2-h^2}]$. The relative phase between the photon and exciton subsystems is then given by (\ref{quadra1}) with the substitution of (\ref{rho_analyt}).

\begin{figure}[t]
\renewcommand{\captionlabeldelim}{.}
\includegraphics[width=\columnwidth]{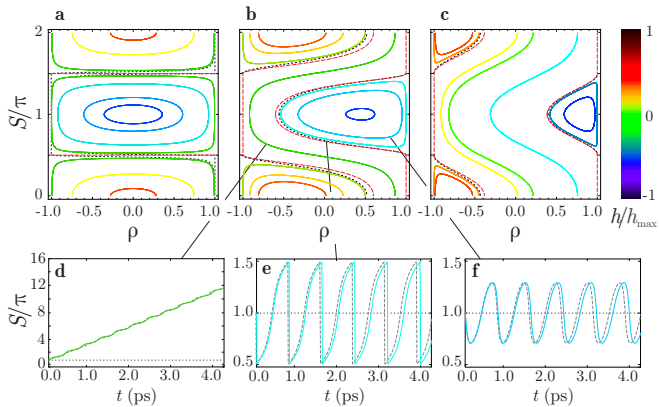}
\caption{\small (Color online) (\textbf{a})--(\textbf{c}) Phase-plane portraits of the conjugate variables $\rho$ and $S$ for different detunings. Trajectories of different colors correspond to different values of the energy $h$ according to the scale given on the right. (\textbf{a}) $\delta=gn/2$, (\textbf{b}) $\delta=-0.5$, (\textbf{c}) $\delta=-1.5$ (all energies are in the units of $\hbar\Omega_R$). For positive detunings, portraits flip with respect to the vertical axis. The red dashed and black dotted lines represent separatrices for particle densities $10^{10}$~cm$^{-2}$ and $0.5\times10^{12}$~cm$^{-2}$, respectively. (\textbf{d})--(\textbf{f}) Relative phase $S=S_C-S_X$ against time, for the trajectories of the phase-plane portrait (\textbf{b}) as marked.  Right to left: (\textbf{f}) anharmonic modification of Rabi oscillations, (\textbf{e}) sawtooth-like oscillations at the separatrix, (\textbf{d}) the regime of running relative phase (internal Josephson effect). For $gn/2=0.002$, analytical expressions (\ref{rho_analyt}), (\ref{quadra1}) coincide with direct numerical solutions of Eqs. (\ref{rho}), (\ref{s}) (the solid lines). The dashed lines show the numerical solutions for $gn/2=0.5\times10^{-1}$. Other physical parameters: $m_C=0.6\times10^{-4} m_0$, $\hbar\Omega_R=5$~meV, $\bar{g}=0.015$~meV$\cdot\mu$m$^2$.}
\label{phase_plane}
\end{figure}

The phase-plane portrait of the conjugate variables $(\rho,S)$ is shown in Fig.~\ref{phase_plane}(a)--(c) for three detuning values. When the detuning compensates the blueshift ($\delta=gn/2$), for any initial conditions the system performs finite motion along the selected trajectory (depending on the energy). This case is displayed in Fig~\ref{phase_plane}(a) and it corresponds to Rabi-like oscillations. Without interactions, if the initial state is prepared in such a way that the exciton and photon populations are equal, \textit{i.e.} $\rho(0)=0$, the system stays in the pure Rabi regime of density oscillations with the time-average $\langle\rho\rangle$ remaining zero in time and without any change of the relative phase $S=\pi$. Any non-zero $\rho(0)$, however, will result in harmonic oscillations in both population imbalance and relative phase with the Rabi frequency $\Omega_R$ ($\omega=1$). Allowing for interactions in this case results in the frequency given by $\omega = 1 + gnh/2$ \cite{elistr}. The obtained correction reveals the decrease of frequency with the amplitude ($h<0$), which is characteristic for nonlinear pendulums, however in this case the frequency does not reduce to zero at the separatrix ($h=0$), tending instead to its non-perturbed value $\Omega_R$.

Fig.~\ref{phase_plane}(b) and \ref{phase_plane}(c) show the phase space trajectories for small and large negative detunings, respectively. In this case, the two-component system can evolve in two different dynamical regimes, depending on the initial energy. Closed trajectories representing finite motion at low energies belong to the regime of Rabi-like oscillations similar to the previous case. The difference consists of the renormalization of the oscillation frequency $\omega$, the anharmonicity of the relative phase oscillations (see Fig.~\ref{phase_plane}(f)), and the shift of the time-average population imbalance to a non-zero value $h\delta/\omega^2$. This regime of oscillations is kind of an interplay between the modified Rabi dynamics and an analog of internal Josephson effect: for small-amplitude oscillations, one may say that the shift of natural frequency corresponds to Josephson ``plasma frequency'' $\omega_{JP}=\delta\Omega_R$.

As the energy $h$ increases at fixed detuning (or, alternatively, as $|\delta|$ increases at fixed $h$), the phase oscillations grow in amplitude up to $\pi/2$ and acquire strongly anharmonic sawtooth profile shown in Fig.~\ref{phase_plane}(e), while the trajectory on the phase-plane portrait approaches the separatrix line defined by $\cos S=\delta\sqrt{(1-\rho)/(1+\rho)}$. After crossing the separatrix, one sees a dramatic change from Rabi-like to Josephson-like dynamics: while the density imbalance oscillates around its new equilibrium value $h\delta/\sqrt{1+\delta^2}$, the relative phase between the photon and exciton condensates $S(t)$ becomes monotonously increasing (or decreasing, depending on the sign of $\delta$) in time like shown in Fig.~\ref{phase_plane}(d). This regime of running phase is analogous to the a.c.~Josephson effect in the Josephson junction \cite{leggett}, or to internal Josephson-like oscillations between the populations in a mixture of spin-up and spin-down atoms when external magnetic field is applied \cite{matthews}. (N.B., all the above explanations imply that the system is that of lower polaritons. For upper polaritons, the relative phase would oscillate around zero instead of $\pi$, and the decrease instead of increase in $h$ would bring the system closer to the separatrix and consequently to the Josephson regime).

Numerical solutions of Eqs.~(\ref{rho}) and (\ref{s}) taking into account interactions start to noticeably differ from the analytical solutions (\ref{rho_analyt}) and (\ref{quadra1}) owhen the dimensionless parameter $gn/2$ becomes of the order of $10^{-1}$ and larger (see Fig.~\ref{phase_plane}(d)--(f)). First-order analytical correction to the separatrix lines appears as $\cos S=(\delta-(1-\rho)gn/4)\sqrt{(1-\rho)/(1+\rho)},$ and is shown in Fig.~\ref{phase_plane}(a)--(c) for $gn/2 = 0.05$ as black dotted lines. The region of finite motion corresponding to modified Rabi oscillations is slightly reduced by interactions, although for higher detunings their influence weakens.

Let us address the preparation of initial states, namely $\rho(0)$ and $S(0)$, which define the energy $h$. When LP and UP branches are resonantly excited by two spectrally narrow, phase-correlated laser pulses, the effective state created in the system is a linear combination of the lower and upper polariton states with controllable populations and relative phase. For example, if the pulses arrive in phase, the initial state will be purely photonic, and if they arrive in antiphase, it will be purely excitonic. The multitude of intermediate cases produce all the variety of possible initial conditions for the considered system.

\begin{figure}[t]
\renewcommand{\captionlabeldelim}{.}
\includegraphics[width=0.7\columnwidth]{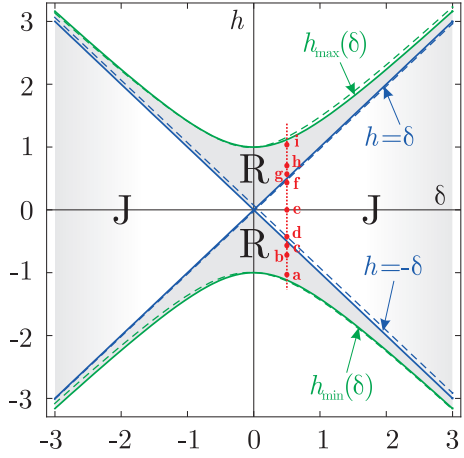}
\caption{\small (Color online) Energy-detuning diagram summarizing the internal dynamics (in the units of $\hbar\Omega_R$). Minimal and maximal energies (per particle) are given by the solid lines $h_{\mbox{\footnotesize min/max}} =\mp\sqrt{1+\delta^2}$ as marked. For LP, $h<0$ and $\langle S\rangle=\pi$, and for UP, $h>0$ and $\langle S\rangle=0$. Within these limits, the different dynamical regimes are divided by the separatrices $h=\pm \delta$. The dashed lines take into account interactions for the polariton density $0.5\times10^{12}$~cm$^{-2}$ and $\bar{g}=0.015$~meV$\cdot\mu$m$^2$. `R'-regions correspond to modified Rabi oscillations. `J'-regions correspond to Josephson-like oscillations with the running relative phase. Red dotted line is a guide to the eye for the corresponding data in Fig.~\ref{fourier}.}
\label{diagram}
\end{figure}
All the dynamical regimes are finally summarized in the energy-detuning diagram displayed in Fig.~\ref{diagram}. In the absence of interactions, dimensionless energy of the LP system (energy per particle) can change in the range $-\omega<h<0$, while for upper polaritons $0<h<\omega$. Corrected by interactions,
\begin{equation}\label{h_int}
h_{\mbox{\footnotesize min/max}}=\mp\,\omega\,\mp\frac{gn}{2}\frac{\delta}{\omega}+\frac{gn}{2} \left(\frac{1}{2}\frac{\delta^2}{\omega^2}+\frac{\delta^2\mp\delta^2}{\omega^4}\right).
\end{equation}
The critical values of $h$ which correspond to the transition between the `modified Rabi' and the `internal Josephson' regimes are defined for each detuning as $h=\pm (\delta-gn/2)+gn/4$. The diagram $h(\delta)$ is divided in four regions which correspond to Rabi-like and Josephson-like oscillations of lower and upper polaritons, as shown in Fig.~\ref{diagram}. The regime of pure Rabi oscillations (with constant relative phase) corresponds to the point $\delta=gn/2$, $h=-1$.
(Or, for the hypothetical equilibrium UP condensate, $\delta=gn/2$, $h=+1$.) It can also be seen from this diagram that the larger is the detuning between the modes, the less extra energy is needed for the transition to the Josephson regime to happen.

\begin{figure}[b]
\renewcommand{\captionlabeldelim}{.}
\includegraphics[width=\columnwidth]{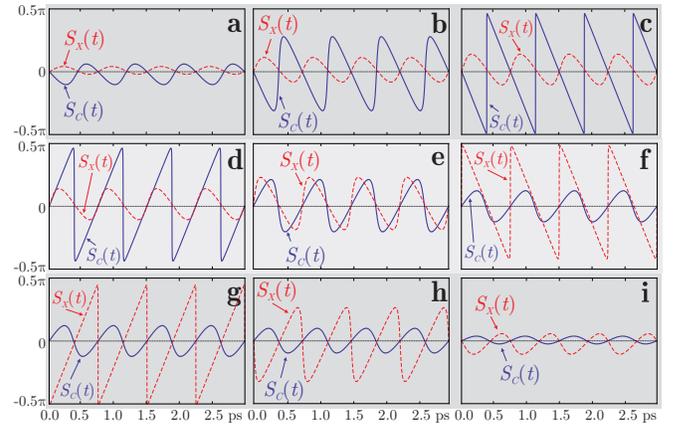}
\caption{\small (Color online) Periodic parts of the photon and exciton phases $S_{C,X}(t)$ as marked, for $\delta=0.5$ and different initial conditions according to the red dotted line in Fig.~\ref{diagram}: (\textbf{a})~$h=0.95h_{\mbox{min}}$, (\textbf{b})~$h=0.6h_{\mbox{min}}$, (\textbf{c})~$h=0.448 h_{\mbox{min}}$, (\textbf{d})~$h=0.44 h_{\mbox{min}}$, (\textbf{e})~$h=0$, (\textbf{f})~$h=0.44h_{\mbox{max}}$, (\textbf{g})~$h=0.448h_{\mbox{max}}$, (\textbf{h})~$h=0.6h_{\mbox{max}}$, (\textbf{i})~$h=0.95h_{\mbox{max}}$. The background color identifies the regimes (`R' or `J') in agreement with Fig.~\ref{diagram}. For negative detunings the curves corresponding to the photon and exciton phases swap. }
\label{fourier}
\end{figure}

Finally, we analyze (for vanishing interactions) how the described internal photon-exciton dynamics influence the phase of the photon field. Using the solutions for $\rho(t)$ and $S(t)$ given by (\ref{rho_analyt}) and (\ref{quadra1}), we analytically integrate Eqs. (\ref{phase_phot}) and (\ref{phase_exc}). The solutions read:

\begin{multline}\label{SC}
S_{C,X}(t) = S_{C,X}(0)-\frac{\epsilon_C^0+\epsilon_X^0}{2}\,t \\
-\arctan\frac{(\omega^2\pm h\delta)\tan\frac{\omega t-\varphi}{2}\pm\sqrt{\omega^2-h^2}}{\omega\,(h\pm\delta)} \\
- \arctan\frac{(\omega^2\pm h\delta)\tan\frac{\varphi}{2}\mp\sqrt{\omega^2-h^2}}{\omega\,(h\pm\delta)}.
\end{multline}
This result highlights the fact that linear rotation of the photon and exciton phases given by $-(\epsilon_C^0+\epsilon_X^0) t/2$ is modulated by additional periodic terms. Fig.~\ref{fourier} shows the periodic parts of the phases (\ref{SC}) for different initial states at fixed detuning $0.5\hbar\Omega_R$. One can clearly see that when the polariton system is in the regions `R' or `J' (as marked in Fig.~\ref{diagram}) and far from the separatrices, the additional periodic terms in both $S_C$ and $S_X$ are approximately harmonic with frequency $\omega$ and of small amplitude (see Fig.~\ref{fourier}(a) and (i)). Upon approaching the separatrix $h=-\delta$, the additional oscillations of the photon phase become strongly anharmonic (Fig.~\ref{fourier}(b)--(e)), and their amplitude grows up to $\pi/2$. Fourier spectrum of the periodic part of $S_C$ then consists of multiple frequencies. Being added to the linear term, these sawtooth oscillations result in the ladder-like behavior of the photon phase. At the same time the evolution of the exciton phase $S_X$ stays practically unchanged. On the contrary, when the energy is close to another critical value $h=+\delta$, the exciton phase experiences the ladder-like behavior while the photon phase oscillations are close to harmonicity (see Fig.~\ref{fourier}(e)--(h)). Experimentally, the photon phase and its evolution can be determined through interferometry \cite{luis} with an external reference with a well-defined phase. Even though the photon phase alone is not enough to retrieve the whole information about the relative phase $S$, it can clearly identify the crossover between the two different dynamical regimes while the initial conditions are being changed.

It is important to note that, when taking into account the out-of-equilibrium nature of polaritons such as their dissipation and replenishing of the system from the reservoir, one expects that the system should show relaxation oscillations to the stable points shown in Fig.\ref{phase_plane}(a)--(c). However, this appears to be the case only only when the gain and loss rates are linear and constant. Modelling the dynamics with non-linearities such as gain saturation \cite{borgh} or the reservoir dynamics \cite{wouters} results in new regimes of evolution, with the detuning and the initial conditions no longer playing a crucial role, therefore they are subject to a separate study.

In conclusion, we analytically analyzed the influence of the photon--exciton energy detuning and repulsive interactions on the internal oscillatory dynamics of the polariton system. We demonstrated that at any non-zero detuning, the two-component system can, depending on its energy, oscillate around its equilibrium point or transit to the regime of monotonously growing relative phase, which we connect to internal Josephon effect.
Interactions, on the contrary, are shown not play a qualitative role in the presented physics. While the Josephson regime we describe is very much analogous to that of conventional bosonic Josephson junctions, the significant difference is that the effects we report lie in the non-interacting regime of the exciton-polariton system. When present, interactions do not significantly modify the dynamics, in contrast to the situation in strong-interacting atomic systems, where interactions could unveil the regime of macroscopic quantum self-trapping of populations \cite{giovanazzi}.
At last, we predict that the crossover between the two regimes of dynamics can be experimentally observed by detecting the photon phase evolution in photoluminescence from the cavity.

Authors thank Alexey Kavokin and Luis Vi\~{n}a for fruitful discussions. This work is partially supported by Russian Foundation for Basic Research. Yu.E.L. is supported by Program of Basic Research of HSE.


\begin{thebibliography}{99}
\bibitem{kir} Densities at which excitons can still be considered bosons can be very roughly estimated as $n_{cr}=1/a_B^2$, $a_B$ being the 2D exciton Bohr radius. For instance, for CdTe and GaAs cavities one has $a_B=3.5$~nm and $15$~nm, respectively, which leads one to critical densities as large as $n_{cr}= 8\times10^{12}$~cm$^{-2}$ and $n_{cr}=4.4\times 10^{11}$~cm$^{-2}$. This density limit must not be reached.
\bibitem{elistr} This result is obtained by means of canonical transformation to action--angle variables in the system without interactions and energy detuning, followed by the use of Poincar\'{e}-Von Zeipel perturbation method to obtain the interaction-induced corrections. Note that this analytical procedure can be performed only for the case $\delta=0$.
\end{thebibliography}
\end{document}